\documentclass[12pt]{article}

\usepackage{amssymb}
\usepackage{amsmath}
\usepackage{cases}
\usepackage{version}
\usepackage{graphicx}

\newtheorem{Prop}{Proposition}[section]

\newtheorem{theor}{Theorem}[section]

\newcommand{\n}{^{(n)}}

\newcommand{\cqfd}{\hfill $\square$}
\newcommand{\R}{\mathbb R}

\newcommand{\varthetab}{{\pmb \vartheta}}

\newcommand{\Deltab}{{\pmb \Delta}}

\newcommand{\taub}{{\pmb \tau}}

\newcommand{\zerob}{{\pmb 0}}

\newcommand{\pr}{^{\prime}}

\title{Efficiency combined with simplicity: new  testing procedures for Generalized Inverse Gaussian models 
}
\author{Angelo Efoevi Koudou \,and \,Christophe Ley \\ {\it \small Universit\'e de Lorraine and CNRS, Institut Elie Cartan de Lorraine}  \\
 {\it \small ECARES and D\' epartement de Math\' ematique, Universit\'{e} libre de Bruxelles} \\ {\it \small Campus Plaine, CP210, Boulevard du Triomphe, B-1050,
Brussels, Belgium}} 
\date{}
\begin{document}
\maketitle
\markboth{ \small Simple efficient tests in GIG models }{ \small Simple efficient tests in GIG models}

\begin{abstract}

The standard efficient testing procedures in the Generalized Inverse Gaussian (GIG) family (also known as Halphen Type A family) are likelihood ratio tests, hence rely on Maximum Likelihood (ML) estimation of the three parameters of the GIG. The particular form of GIG densities, involving modified Bessel functions, prevents in general from a closed-form expression for ML estimators, which are obtained at the expense of complex numerical approximation methods. On the contrary, Method of Moments (MM) estimators allow for concise expressions, but tests based on these estimators suffer from a lack of efficiency compared to likelihood ratio tests. This is why, in recent years, trade-offs between ML and MM estimators have been proposed, resulting in simpler yet not completely efficient estimators and tests. In the present paper, we do not propose such a trade-off but rather an optimal combination of both methods, our tests inheriting efficiency from an ML-like construction  and simplicity from the MM estimators of the nuisance parameters. This goal shall be reached by attacking the problem from a new angle, namely via the Le Cam methodology. Besides providing simple efficient testing methods, the theoretical background of this methodology further allows us to write out explicitly power expressions for our tests. A Monte Carlo simulation study shows that, also at small sample sizes, our simpler procedures do at least as good as the complex likelihood ratio tests. We conclude the paper by applying our findings on two real-data sets.
\end{abstract}

{\it Keywords:} 
Asymptotic linearity, GIG distributions, IG distributions, Maximin tests, Uniform local asymptotic normality

{\it AMS 2000 subject classifications}: Primary 62F03; Secondary 62F05

\section{Introduction.}

The \emph{generalized inverse Gaussian} (hereafter GIG) distribution with parameters $p\in\R$, $a>0$, $b>0$  has density
\begin{equation}\label{GIG}
f_{p,a,b}(x):=c(p,a,b) x^{p-1}e^{-(ax+b/x)/2}, \quad x>0,
\end{equation}
with normalizing constant 
$$
c(p,a,b):=\frac{(a/b)^{p/2}}{2K_p(\sqrt{ab})}
$$
where $K_p$ is the modified Bessel function of the third kind. The parameters $a$ and $b$ regulate both the concentration and scaling of the densities, the former via  $\sqrt{ab}$ and the latter via $\sqrt{b/a}$, while the parameter $p$ bears no precise statistical meaning. This is why some authors rather use the parameterization $\theta=\sqrt{ab}$ and $\eta=\sqrt{b/a}$, but we stick here to $a$ and $b$. The GIG family has been proposed in Good (1953), although Halphen~(1941) had already previously discussed such distributions, which is why in some articles one speaks of Halphen Type A distributions instead of GIG distributions (see Seshadri~1999). GIG distributions enjoy several nice probabilistic features such as, e.g., the equivalence $X \sim {\rm GIG}(p , a, b)\Longleftrightarrow \frac{1}{X} \sim {\rm GIG}(-p , b, a)$ (see Barndorff-Nielsen and Halgreen~1977, where some convolution properties and infinite divisibility are pointed out). These distributions have been used in the modelization of diverse phenomena such as, for instance, waiting time (J\o rgensen~1982), neural activity (Iyengar and Liao~1997),  or, most importantly, hydrologic extreme events (see Chebana \emph{et al.}~2010 and references therein). The GIG family also contains several well-known sub-models such as the Gamma distribution (for $b=0$ and $p>0$), the reciprocal Gamma distribution (for $a=0$ and $p<0$), the hyperbolic distribution (for $p=0$), the reciprocal inverse Gaussian (hereafter RIG) distribution (for $p=\frac{1}{2}$) and the \emph{Inverse Gaussian} (hereafter IG) distribution (for $p=-\frac{1}{2}$).

While numerous papers address probabilistic aspects of GIG distributions, relatively few articles discuss their statistical properties. The cornerstone reference in this respect is J\o rgensen~(1982, Chapter 4-7), complemented by the stream of literature on general Halphen distributions (of which the GIG or Halphen A is one of three representatives), see e.g. Perreault \emph{et al.}~(1999a,b) or Chebana \emph{et al.}~(2010). 
Regarding   hypothesis testing, the standard tools are likelihood ratio tests based on maximum likelihood (ML) estimation of the parameters both under the null and without constraints. Despite being the most efficient procedures, ML-based methods suffer from computational complexity as in general no closed-form solutions exist for the likelihood equations in the GIG case (especially when all three parameters are estimated jointly). Improved numerical methods for ML estimation have been put forward by, e.g., Perreault \emph{et al.}~(1999b) and Lemonte and Cordeiro~(2011). As alternative to the complex ML estimation, Fitzgerald~(2000) has given exact expressions for method of moments (MM) estimators, which are thus analytically simple, but tests based on them suffer from a lack of efficiency compared to likelihood ratio tests. In order to find a trade-off between efficiency and simplicity, Chebana \emph{et al.}~(2010) propose mixtures of ML and MM estimation.

In the present paper, our aim is not a such a trade-off but rather an optimal combination of both approaches, the resulting tests inheriting efficiency from an ML-like construction and simplicity from  MM estimation of the nuisance parameters under the null (no further non-null estimation will be required). We shall achieve this by having recourse to the \emph{Le Cam methodology}, whose first step consists in showing that GIG distributions satisfy the \emph{Uniform Local Asymptotic Normality (ULAN)} property. With this key property in hand, we build  optimal test procedures for any null hypothesis involving one or more of the parameters $p, a$ and $b$, the nuisance parameters remaining unspecified. The resulting test statistics resemble Rao score (or Lagrange Multiplier) statistics, hence are as efficient as the ML-based likelihood ratio tests, but improve on the Rao score tests by the fact that the nuisance parameters need not be ML-estimated under the null, but can be exactly calculated as MM-estimators. 
Yet another attractive feature of the Le Cam approach consists in the fact that we are able to calculate the power of our tests against sequences of contiguous alternatives.

The paper is organized as follows. In Section~\ref{sec:ULAN}, we  derive the  ULAN property for  ${\rm GIG}(p,a,b)$ models and the ensuing crucial \emph{asymptotic linearity}. In Section \ref{sec:test} we show how to construct  optimal  tests for the  null hypothesis $\mathcal{H}_0:p=p_0$ against $\mathcal{H}_1^{\neq}:p\neq p_0$, determine their asymptotic behavior both under the null and under a sequence of contiguous alternatives and write out explicitly the asymptotic powers.  In Section~\ref{sec:IG}, we then particularize our findings to $p_0=-1/2$, corresponding to goodness-of-fit tests for the IG distribution within GIG models (which yields for free a goodness-of-fit testing procedure for RIG distributions as $X\sim{\rm IG}(a,b)\Longleftrightarrow \frac{1}{X}\sim{\rm RIG}(b,a)$). A Monte Carlo simulation study allows us to investigate the   finite-sample behavior of our tests.  We then use, in Section~\ref{sec:real}, our procedures  in order to analyze two real-data examples. A brief outlook on future research is provided in Section~\ref{sec:future}. Finally, an Appendix collects the technical proofs.

\section{The ULAN property and asymptotic linearity of general ${\rm GIG}(p,a,b)$ models.}\label{sec:ULAN}

Let $X_1,\ldots,X_n$ be \mbox{i.i.d.} observations following the ${\rm GIG}(p,a,b)$ distribution with density~(\ref{GIG}), and consider the GIG-parametric model 
$$\mathcal{P}^{(n)}_{GIG}:=\left\{{\rm P}^{(n)}_{p,a,b}:p\in\R,a\in\R_0^+,b\in\R_0^+\right\},$$
where ${\rm P}^{(n)}_{p,a,b}$ stands for the joint distribution of $(X_1,\ldots,X_n)$.  The ULAN property of the ${\rm GIG}(p,a,b)$ model is achieved in the following result, whose proof is given in the Appendix.



\begin{theor}\label{ULAN}
Let $\varthetab=(p,a,b)'$ for any $p\in\R, a\in\R^+_0$ and $b\in\R^+_0$. Then, for perturbation rates of order $n^{-1/2}$, the family of probability distributions $\mathcal{P}^{(n)}_{GIG}$ is ULAN at $\varthetab$ with central sequence
\begin{eqnarray*}
\Deltab^{(n)}(\varthetab)
:=
\left( 
\begin{array}{c}
\Delta^{(n)}_{p}(\varthetab)\\[1mm]
\Delta^{(n)}_{a}(\varthetab)\\[1mm]
\Delta^{(n)}_{b}(\varthetab)
\end{array}
\right)
&:=&
\frac{1}{\sqrt{n}}\, \sum_{i=1}^n 
\left( 
\begin{array}{c}
\frac{\partial_p c(p,a,b)}{c(p,a,b)}+\log(X_i)\\
\frac{\partial_a c(p,a,b)}{c(p,a,b)}-\frac{X_i}{2}\\
\frac{\partial_b c(p,a,b)}{c(p,a,b)}-\frac{1}{2X_i}\\
\end{array}
\right) \nonumber \\
& = & 
\frac{1}{\sqrt{n}}\, \sum_{i=1}^n 
\left( 
\begin{array}{c}
   \frac{1}{2} \log (a/b) -\partial_p \log K_p(\sqrt{ab})        +\log(X_i)\\
\frac{p}{2a}   -\partial_a \log K_p(\sqrt{ab})   -\frac{X_i}{2}\\
-\frac{p}{2b}  -\partial_b \log K_p(\sqrt{ab})  -\frac{1}{2X_i}\\
\end{array}
\right) \nonumber \\
\end{eqnarray*}
and corresponding Fisher information matrix 
$${\it \Gamma}(\varthetab):=\left(
\begin{array}{ccc}
\Gamma_{p,p}(\varthetab)&\Gamma_{p,a}(\varthetab)&\Gamma_{p,b}(\varthetab)\\
\Gamma_{p,a}(\varthetab)&\Gamma_{a,a}(\varthetab)&\Gamma_{a,b}(\varthetab)\\
\Gamma_{p,b}(\varthetab)&\Gamma_{a,b}(\varthetab)&\Gamma_{b,b}(\varthetab)\\
\end{array}\right),$$
where 
$$\Gamma_{p,p}(\varthetab):= \partial_{pp}^2 \log K_p(\sqrt{ab}), \quad \Gamma_{a,a}(\varthetab):=  \frac{p}{2a^2} + \partial_{aa}^2 \log K_p(\sqrt{ab})  ,$$
$$\Gamma_{b,b}(\varthetab):= -\frac{p}{2b^2} + \partial_{bb}^2 \log K_p(\sqrt{ab})  , \quad \Gamma_{p,a}(\varthetab):= -\frac{1}{2a} + \partial_{ap}^2 \log K_p(\sqrt{ab})  ,$$
$$\Gamma_{p,b}(\varthetab):=  \frac{1}{2b} + \partial_{bp}^2 \log K_p(\sqrt{ab})\quad \mbox{and} \quad
\Gamma_{a,b}(\varthetab):=   \partial_{ab}^2 \log K_p(\sqrt{ab})    .$$

More precisely, for any $\varthetab^{(n)}=(p^{(n)},a^{(n)}, b^{(n)})'=\varthetab+O(n^{-1/2})$ and for any bounded sequence $\taub^{(n)}=(\tau_1^{(n)},\tau_2^{(n)},\tau_3^{(n)})'\in\R^3$ such that $a^{(n)}+n^{-1/2}\tau_2\n>0$ and $b^{(n)}+n^{-1/2}\tau_3\n>0$, we have
\begin{eqnarray}
\Lambda^{(n)}_{\varthetab^{(n)}+n^{-1/2}\taub^{(n)}/\varthetab^{(n)}}&:=&\log(d{\rm P}^{(n)}_{\varthetab^{(n)}+n^{-1/2}\taub^{(n)}}/d{\rm P}^{(n)}_{\varthetab^{(n)}})\nonumber\\
&=&(\taub^{(n)})'\Deltab^{(n)}(\varthetab^{(n)})-\frac{1}{2}(\taub^{(n)})'{\it \Gamma}(\varthetab)\taub^{(n)}+o_{\rm P}(1)\label{Taylor}
\end{eqnarray}
and $\Deltab^{(n)}(\varthetab^{(n)})\stackrel{\mathcal{L}}{\rightarrow}\mathcal{N}_3(\zerob,{\it \Gamma}(\varthetab))$, both under ${\rm P}^{(n)}_{\varthetab^{(n)}}$ as $n\rightarrow\infty$.
\end{theor}\vspace{2mm}

The central idea of the Le Cam theory is the concept of convergence of statistical models (\emph{experiments} in the Le Cam vocabulary). Quoting Le Cam~(1960), ``the family of probability measures under study can be approximated very closely by a family of a simpler nature''. The key ingredient in this approximation is the ULAN property, from which we can deduce that (see Le Cam and Yang~2000, \mbox{page 89} for details) our GIG-parametric model $\mathcal{P}^{(n)}_{GIG}$ is locally (around $(p,a,b)\pr$) and asymptotically (for large sample sizes) equivalent to a simple Gaussian shift model. Intuitively, this is due to the fact that the likelihood ratio expansion~(\ref{Taylor}), up to the remainder terms, looks like the likelihood ratio of a Gaussian shift model
$$\mathcal{P}_{\pmb{\vartheta}}:=\left\{ \rm{P}_{ {\taub},  {\varthetab}}=\mathcal{N}_3\left({\it \Gamma}(\varthetab) {\taub}, {\it \Gamma}(\varthetab)\right)| \,{\taub} \in {\R}^3\right\}$$
with a single observation which we denote as $\pmb{\Delta}$. This means that all power functions that are implementable in the local GIG experiments are the power functions that are possible in the Gaussian shift
experiment. In view of these considerations, it follows that asymptotically optimal
tests in our local models can be derived by analyzing the Gaussian limit model, for which the most efficient procedures are well-known. The detailed construction is described in the next section. We conclude the present section by writing down an immediate consequence of the ULAN property, namely the following \textit{asymptotic linearity} property of the central sequence:
\begin{align}\label{aslin} {\Deltab}^{(n)}\left(\varthetab+n^{-1/2} {\taub}^{(n)}\right)={\Deltab}^{(n)}(\varthetab)-{\it \Gamma}(\varthetab)\taub^{(n)}+o_{\rm{P}}(1),\end{align}
as $n\to\infty$, under ${\rm P}^{(n)}_{\varthetab^{(n)}}$. This asymptotic linearity, combined with the ULAN property, forms the basis for  our hypothesis test statistics, as it reveals us the behavior of the central sequence when $\varthetab=(p,a,b)$ is replaced with some estimator $\hat{\varthetab}^{(n)}$. As we shall show in details in the next section, for root-$n$ consistent estimators satisfying some mild regularity assumption, we can replace $\taub^{(n)}$ in~\eqref{aslin} with $n^{1/2}(\hat{\varthetab}^{(n)}-\varthetab)$, yielding
$$
{\Deltab}^{(n)}\left(\hat{\varthetab}^{(n)}\right)={\Deltab}^{(n)}(\varthetab)-{\it \Gamma}(\varthetab)n^{1/2}(\hat{\varthetab}^{(n)}-\varthetab)+o_{\rm{P}}(1)
$$
as $n\to\infty$ under ${\rm P}^{(n)}_{\varthetab^{(n)}}$. This nice asymptotic equality will be put to use in what follows.

\section{Construction of efficient testing procedures.}\label{sec:test}

Based on our findings of the previous section, we shall describe in the present section how to tackle hypothesis testing problems in a way as efficient as likelihood ratio tests but with test statistics whose expressions can be  explicitly written down, taking advantage of the exact MM expressions. We focus here on the construction of testing procedures for the  null hypothesis $\mathcal{H}_0:p=p_0$  against $\mathcal{H}_1^{\neq}:p\neq p_0$ for  $p_0\in\R$. This will be done in two steps: first, we assume that the nuisance parameters (here $a$ and $b$) are known, and second, we consider them as unknown and hence they need to be estimated. Of course, the construction below is by no means restricted to the parameter $p$ and can be used for hypothesis tests about $a$ and $b$, about any of the vectors $(p,a)\pr$, $(p,b)\pr$ and $(a,b)\pr$, and even about the vector $(p,a,b)\pr$. The latter case is evidently the simplest, as the aforementioned second step will not be needed. \vspace{0.5cm}

\noindent\underline{Step 1: the nuisance parameters $a$ and $b$ are known}\vspace{3mm}

\noindent In this scenario, it suffices to read the ULAN property only in the parameter $p$, and no asymptotic linearity is needed. By analogy with the Gaussian shift experiment, the optimal testing procedure for $p=p_0$ against $p\neq p_0$  consists in rejecting the null at asymptotic level $\alpha$ whenever the absolute value of the test statistic
\begin{eqnarray*}
Q\n_{p_0}(a,b)&:=&\frac{\Delta^{(n)}_{p}(p_0,a,b)}{\sqrt{\Gamma_{p,p}(p_0,a,b)}}\nonumber\\
&=&\frac{n^{-1/2}\sum_{i=1}^n\left(\frac{1}{2} \log (a/b) -(\partial_p \log K_p(\sqrt{ab}))|_{p=p_0} +\log(X_i)\right)}{\sqrt{(\partial_{pp}^2 \log K_p(\sqrt{ab}))|_{p=p_0}}}
\end{eqnarray*}
exceeds the $\alpha/2$-upper quantile $z_{\alpha/2}$ of a standard Gaussian distribution.
We give at the end of the current section simplified forms for $\partial_p \log K_p(\sqrt{ab}) $ and 
$\partial_{pp}^2 \log K_p(\sqrt{ab})$. The validity of this test, which we denote $\phi^{(n)}_{p_0}(a,b)$, and its optimality follow from the more general result in Theorem~\ref{behav} below.\vspace{0.5cm}

\noindent\underline{Step 2: the nuisance parameters $a$ and $b$ are unknown}\vspace{3mm}

\noindent This situation is clearly the less specific and more realistic one, but of course also more complicated. Indeed, since both $a$ and $b$ are unknown, we need to estimate them and plug the corresponding estimators $\hat a\n$ and $\hat b\n$ into the test statistic $Q\n_{p_0}(a,b)$ to yield $Q\n_{p_0}(\hat a\n,\hat b\n)$. However, such a replacement cannot be achieved without care, and its asymptotic effects need to be calculated. A first observation can be made by looking at the information matrix ${\it \Gamma}(\varthetab)$. Would ${\it \Gamma}(\varthetab)$ be block-diagonal (in the sense that no correlation exists between the blocks $p$ and $(a,b)\pr$), then the substitution of $(\hat{a}\n,\hat{b}\n)\pr$ for $(a,b)\pr$ would have no influence, asymptotically, on the behavior of the central sequence $\Delta^{(n)}_{p}(\varthetab)$. However, this is obviously not the case here, and consequently a local perturbation of $a$ or $b$ has the same asymptotic impact on $\Delta^{(n)}_{p}(\varthetab)$ as a local perturbation of $p$ around $p_0$. It follows that the cost of not knowing the true values of the nuisance parameters $a$ and $b$  is strictly positive; the stronger the correlation between the central sequences in $p$ and $(a,b)\pr$, the larger that cost. 

From this short discussion on the information matrix, it becomes clear that a more desirable central sequence for $p$ would enjoy this block-diagonality, in other words, we need a ``decorrelated'' central sequence. This decorrelation takes into account the aforementioned cost of not knowing $a$ and $b$ and is achieved by the so-called \emph{efficient central sequence} for $p$
\begin{eqnarray*}
&&\Delta^{(n)eff}_{p}(\varthetab)\nonumber\\
&:=&\Delta^{(n)}_{p}(\varthetab)-{\rm Cov}(\Delta^{(n)}_{p}(\varthetab),(\Delta^{(n)}_{a}(\varthetab),\Delta^{(n)}_{b}(\varthetab)))\left({\rm Var}((\Delta^{(n)}_{a}(\varthetab),\Delta^{(n)}_{b}(\varthetab))\pr)\right)^{-1}(\Delta^{(n)}_{a}(\varthetab),\Delta^{(n)}_{b}(\varthetab))\pr \nonumber\\
&=&\Delta^{(n)}_{p}(\varthetab)-\frac{(\Gamma_{p,a}(\varthetab),\Gamma_{p,b}(\varthetab))\left(\begin{array}{cc}
\Gamma_{b,b}(\varthetab)&-\Gamma_{a,b}(\varthetab)\\
-\Gamma_{a,b}(\varthetab)&\Gamma_{a,a}(\varthetab)
\end{array}\right)
(\Delta^{(n)}_{a}(\varthetab),\Delta^{(n)}_{b}(\varthetab))\pr}{\Gamma_{a,a}(\varthetab)\Gamma_{b,b}(\varthetab)-(\Gamma_{a,b}(\varthetab))^2}.\nonumber\\
\end{eqnarray*}
This efficient central sequence is obtained by projecting $\Delta^{(n)}_{p}(\varthetab)$ onto the subspace orthogonal to $(\Delta^{(n)}_{a}(\varthetab),\Delta^{(n)}_{b}(\varthetab))\pr$, which ensures that the new efficient central sequence $\Delta^{(n)eff}_{p}(\varthetab)$ is asymptotically uncorrelated with the central sequences corresponding to $a$ and $b$. The Fisher information quantity associated with $\Delta^{(n)eff}_{p}(\varthetab)$ is given by
\begin{eqnarray*}
\Gamma^{eff}_{p,p}(\varthetab)&:=&\Gamma_{p,p}(\varthetab)-\frac{(\Gamma_{p,a}(\varthetab),\Gamma_{p,b}(\varthetab))\left(\begin{array}{cc}
\Gamma_{b,b}(\varthetab)&-\Gamma_{a,b}(\varthetab)\\
-\Gamma_{a,b}(\varthetab)&\Gamma_{a,a}(\varthetab)
\end{array}\right)
(\Gamma_{p,a}(\varthetab),\Gamma_{p,b}(\varthetab))\pr}{\Gamma_{a,a}(\varthetab)\Gamma_{b,b}(\varthetab)-(\Gamma_{a,b}(\varthetab))^2}\nonumber\\
&=&\frac{\det\Gamma (\varthetab) }{\Gamma_{a,a}(\varthetab)\Gamma_{b,b}(\varthetab)-(\Gamma_{a,b}(\varthetab))^2}.\nonumber
\end{eqnarray*}
Now that we have derived this decorrelated efficient central sequence and its information matrix, it becomes of interest to establish their asymptotic linearity. This is achieved in the following Proposition, whose proof heavily relies on the asymptotic linearity~(\ref{aslin}).

\begin{Prop}\label{aslinprop}
For any  $\varthetab=(p,a,b)'\in\R\times(\R_0^+)^2$, and for any bounded sequence ${\taub}^{(n)}=(0,\tau_2^{(n)},\tau_3^{(n)})' \in \{0\}\times\mathbb{R}^2$ such that $a+n^{-1/2}{\tau_2}^{(n)}>0$ and $b+n^{-1/2}{\tau_3}^{(n)}>0$, we have that, under ${\rm P}_{\varthetab}^{(n)}$ and as~$n\rightarrow\infty$,
\begin{equation}\label{aslineff}
\Delta^{(n)eff}_{p}(\varthetab+n^{-1/2}\taub^{(n)})=\Delta_p^{(n)eff}(\varthetab)+o_{\rm{P}}(1)
\end{equation}
and
\begin{equation}\label{inflin}
\Gamma^{eff}_{p,p}(\varthetab+n^{-1/2}\taub^{(n)})=\Gamma^{eff}_{p,p}(\varthetab)+o_{\rm P}(1).
\end{equation}
\end{Prop}

The proof is deferred to the Appendix. 
As explained at the end of the previous section, the idea behind this asymptotic linearity result consists in using respectively $a+n^{-1/2}\tau_2\n=\hat{a}\n$ and $b+n^{-1/2}\tau_3\n=\hat{b}\n$, that is, $\tau_2\n=n^{1/2}(\hat{a}\n-a)$ and $\tau_3\n=n^{1/2}(\hat{b}\n-b)$. Both sequences remain bounded if the estimators $\hat{a}\n$ and $\hat{b}\n$ are root-$n$ consistent, a very natural requirement. However, in order to perform this replacement (which is not evident as the non-random sequences $\tau_1^{(n)}$ and $\tau_2^{(n)}$ are replaced with random sequences), one more condition needs to be imposed on the estimators, and this is summarized in the following \vspace{3mm}

{\sc  Assumption~A}. 
\emph{
The sequence of estimators $\hat a\n$ and $\hat b\n$ is \textrm{(i)} root-$n$ consistent 
(i.e., $n^{1/2}(\hat a\n-a)=O_{\rm P}(1)$ and $n^{1/2}(\hat b\n-b)=O_{\rm P}(1)$ as $n\rightarrow\infty$, under  ${\rm P}\n_{p,a,b}$) and \textrm{(ii)} locally asymptotically discrete,
meaning that, for all $a,b\in\R_0^+$ and all $c>0$, there exists an $M=M(c)>0$ such that the number of possible values of $\hat a\n$ and $\hat b\n$ in intervals of the form $\{t\in\R : n^{1/2}|t-a| \leq c\}$ and $\{t\in\R : n^{1/2}|t-b| \leq c\}$ is bounded by $M$, uniformly as $n\rightarrow \infty$.
}
\vspace{4mm}

It should be noted that Assumption~A(ii) is a purely technical requirement, with little practical implications (for fixed sample size, any estimator indeed can be considered part of a locally asymptotically discrete sequence; see Le Cam and Yang~2000). This condition is however essential as it precisely allows to replace $\tau_2\n$ with $n^{1/2}(\hat{a}\n-a)$ and $\tau_3\n$ with $n^{1/2}(\hat{b}\n-b)$ in~(\ref{aslineff}) and (\ref{inflin}) by Lemma~4.4 in Kreiss~(1987), where such replacements have been theoretically worked out in detail. Using this fact in combination with Slutsky's Lemma, we have the following crucial result.


\begin{Prop}\label{propo}
Let Assumption~A hold and fix $p_0\in\R$. Then, defining 
$$
Q^{(n)eff}_{p_0}(a,b)=\frac{\Delta^{(n)eff}_{p}(p_0,a,b)}{\sqrt{\Gamma^{eff}_{p,p}(p_0,a,b)}},
$$
we have that
\begin{equation}\label{replace}
Q^{(n)eff}_{p_0}(\hat a\n,\hat b\n)-Q^{(n)eff}_{p_0}(a,b)=o_{\rm P}(1)
\end{equation}
under ${\rm P}^{(n)}_{p_0,a,b}$ as $n\rightarrow\infty$.
\end{Prop}
This  finally enables us to derive the optimal testing procedure for $\mathcal{H}_0:p=p_0$ against $\mathcal{H}_1^{\neq}:p\neq p_0$ when the nuisance parameters $a$ and $b$ are unknown. This test, which we denote $\phi^{(n)}_{p_0}(\hat a\n,\hat b\n)$, rejects the null hypothesis at asymptotic level $\alpha$ whenever the absolute value of the test statistic $Q^{(n)eff}_{p_0}(\hat a\n,\hat b\n)$ exceeds $z_{\alpha/2}$, with $\hat{a}^{(n)}$ and $\hat{b}^{(n)}$ satisfying Assumption~A. In the next theorem, we formally establish the validity of this test (behavior of $Q^{(n)eff}_{p_0}(\hat a\n,\hat b\n)$ under the null), its asymptotic distribution under a sequence of local alternatives and its optimality features; of course, the results also cover the case of the easier test $\phi^{(n)}_{p_0}(a,b)$.

\begin{theor}\label{behav}
Fix $p_0\in\R$ and suppose that $\hat{a}^{(n)}$ and $\hat{b}^{(n)}$ satisfy Assumption~A. Then \begin{itemize}
\item[(i)] $Q^{(n)eff}_{p_0}(\hat a\n,\hat b\n)$ is asymptotically standard normal under $\bigcup_{a \in \R_0^+} \bigcup_{b\in\R_0^+}{\rm P}^{(n)}_{p_0,a,b}$;
\item[(ii)] for any $a\in\R_0^+$ and any $b\in\R_0^+$, $Q^{(n)eff}_{p_0}(\hat a\n,\hat b\n)$ is asymptotically  normal with~mean $\tau_1\sqrt{\Gamma^{eff}_{p,p}(p_0,a,b)}$~and variance 1 under ${\rm P}^{(n)}_{p_0+n^{-1/2}\tau_1^{(n)},a,b}$, where  $\tau_1^{(n)} \in \R$ is a bounded sequence and $\tau_1:=\lim_{n \to \infty} \tau_1^{(n)}$;
\item[(iii)] the  test $\phi^{(n)}_{p_0}(\hat a\n,\hat b\n)$, which rejects the null hypothesis $\mathcal{H}_0:=\bigcup_{a \in \R_0^+} \bigcup_{b\in\R_0^+}{\rm P}^{(n)}_{p_0,a,b}$ whenever $|Q^{(n)eff}_{p_0}(\hat a\n,\hat b\n)|>z_{\alpha/2}$, has asymptotic level $\alpha$ under $\mathcal{H}_0$ and is locally and asymptotically maximin for testing $\mathcal{H}_0:p=p_0$ against  $\mathcal{H}_1^{\neq}:=\bigcup_{a \in \R_0^+} \bigcup_{b\in\R_0^+}\bigcup_{p\neq p_0}{\rm P}^{(n)}_{p,a,b}$.
\end{itemize}
\end{theor}

Part~(iii) of Theorem~\ref{behav} provides the theoretical proof of efficiency of our test $\phi^{(n)}_{p_0}(\hat a\n,\hat b\n)$, which is thus as efficient as the likelihood ratio test $\phi^{\rm LR}_{p_0}$. However, the advantage of $\phi^{(n)}_{p_0}(\hat a\n,\hat b\n)$ over  $\phi^{\rm LR}_{p_0}$ lies in its simplicity, as no ML estimation is required here since any root-$n$ consistent estimators for $a$ and $b$ can be used. Hence we opt for their MM estimators, whose expressions, due to Fitzgerald~(2000), are given by $\hat{a}^{(n)}_{\rm MM}:=\frac{\hat{\theta}^{(n)}_{\rm MM}}{\hat{\eta}^{(n)}_{\rm MM}}$ and $\hat{b}^{(n)}_{\rm MM}:= \hat{\theta}^{(n)}_{\rm MM}\hat{\eta}^{(n)}_{\rm MM}$ where
$$
\hat{\eta}^{(n)}_{\rm MM}:=\sqrt{\frac{\bar{X}_{-1}s^2-\bar{X}(\bar{X}\bar{X}_{-1}-1)}{\bar{X}(s_{-1})^2-\bar{X}_{-1}(\bar{X}\bar{X}_{-1}-1)}}
$$
and
$$
\hat{\theta}^{(n)}_{\rm MM}:=\frac{2\left(\frac{1}{\hat{\eta}^{(n)}_{\rm MM}}\bar{X}-\hat{\eta}^{(n)}_{\rm MM}\bar{X}_{-1}\right)}{\frac{1}{(\hat{\eta}^{(n)}_{\rm MM})^2}s^2-(\hat{\eta}^{(n)}_{\rm MM})^2(s_{-1})^2}
$$
with $\bar{X} = \frac{1}{n}\sum_{i=1}^n X_i$, $s^2=\frac{1}{n}\sum_{i=1}^n(X_i-\bar{X})^2$, $\bar{X}_{-1} = \frac{1}{n}\sum_{i=1}^n \frac{1}{X_i}$ and $(s_{-1})^2=\frac{1}{n}\sum_{i=1}^n(\frac{1}{X_i}-\bar{X}_{-1})^2$. Thus, \emph{in fine}, the computationally simple yet efficient test we propose is $\phi^{(n)}_{p_0}:=\phi^{(n)}_{p_0}(\hat a\n_{\rm MM},\hat b\n_{\rm MM})$ which rejects the null hypothesis at asymptotic level~$\alpha$ whenever the absolute value of
$$
Q\n_{p_0}:=Q^{(n)eff}_{p_0}(\hat a\n_{\rm MM},\hat b\n_{\rm MM})=\frac{\Delta^{(n)eff}_{p}(p_0,\hat a\n_{\rm MM},\hat b\n_{\rm MM})}{\sqrt{\Gamma^{eff}_{p,p}(p_0,\hat a\n_{\rm MM},\hat b\n_{\rm MM})}}
$$
exceeds $z_{\alpha/2}$. One-sided tests of the form $\mathcal{H}_0:p=p_0$ against $\mathcal{H}_1^>:p>p_0$ or $\mathcal{H}_1^<:p<p_0$ are of course readily constructed along the same lines and their asymptotic behavior is as well regulated by Theorem~\ref{behav}.

As  mentioned in the Introduction, a welcomed feature of the Le Cam approach lies in the fact that it enables us to derive explicit power expressions under sequences of  local alternatives of the form ${\rm P}^{(n)}_{p_0+n^{-1/2}\tau_1^{(n)},a,b}$ for any $a,b\in\R_0^+$, with $\tau_1^{(n)}$ a bounded sequence with limit $\tau_1$. These powers are easily determined via  Part~(ii) of Theorem~\ref{behav}, as the latter provides us with the asymptotic distribution of $Q^{(n)eff}_{p_0}(\hat a\n,\hat b\n)$ under that sequence of local alternatives.  Denoting by $\Phi$ the cumulative distribution function of the standard Gaussian distribution, the asymptotic power of  $\phi^{(n)eff}_{p_0}(\hat a\n,\hat b\n)$ is  given by
$$1-\Phi\left(z_{\alpha/2}-\tau_1\sqrt{\Gamma_{p,p}^{eff}(p_0,a,b)}\right)+\Phi\left(-z_{\alpha/2}-\tau_1\sqrt{\Gamma_{p,p}^{eff}(p_0,a,b)}\right),$$
and by 
$$1-\Phi\left(z_\alpha-\tau_1\sqrt{\Gamma_{p,p}^{eff}(p_0,a,b)}\right) \qquad \textnormal{and}\qquad \Phi\left(-z_\alpha-\tau_1\sqrt{\Gamma_{p,p}^{eff}(p_0,a,b)}\right)$$
in the respective one-sided tests  against $\mathcal{H}_1^{>}: p>p_0$ and $\mathcal{H}_1^{<}: p <p_0$.

We conclude this section by providing more user-friendly expressions for the entries of the Fisher information matrix ${\it \Gamma} (\varthetab)$. For $j$ taking values in $\{p-3,p-2,p-1,p,p+1\}$, consider the integrals \begin{eqnarray*}
I_j &:=&  \int_0^\infty u^p \exp \left(-\frac{1}{2}(au+b/u)\right)\,du, \\
M_j& := & \int_0^\infty u^p \log(u) \exp \left(-\frac{1}{2}(au+b/u)\right)\,du, \\
 N_j &:= & \int_0^\infty u^{p-1} (\log (u))^2 \exp \left(-\frac{1}{2}(au+b/u)\right)\,du. 
\end{eqnarray*}
For the sake of readability, we omit writing $I_j(a,b), \, M_j(a,b)$ and $N_j(a,b)$ and tacitly assume that each integral depends on $a$ and $b$. The entries of the matrix ${\it \Gamma} (\varthetab)$ can be written as 
$$
\Gamma_{p,p}(\varthetab)=\frac{N_p}{I_{p-1}}-\left(\frac{M_{p-1}}{I_{p-1}} \right)^2, \,\,
\Gamma_{p,a}(\varthetab) = -\frac{1}{2} \left( \frac{M_p}{I_{p-1}}- \frac{I_p M_{p-1}}{I_{p-1}^2} \right),$$
$$\Gamma_{p,b}(\varthetab) = -\frac{1}{2} \left( \frac{M_{p-2}}{I_{p-1}}- \frac{M_{p-1}I_{p-2}}{I_{p-1}^2} \right), \, \,
\Gamma_{a,b}(\varthetab) =  \frac{1}{4} \left( 1- \frac{I_pI_{p-2}}{I_{p-1}^2} \right),
$$
$$
\Gamma_{a,a}(\varthetab)=\frac{1}{4} \left( I_{p+1}- I_p^2 \right),\,\, \mbox{and}\,\, \Gamma_{b,b}(\varthetab)=\frac{1}{4} \left( I_{p-3}- I_{p-2}^2 \right). 
$$
This way of writing the information matrix renders its calculation easier in programs such as $\mathtt{R}$ or MATHEMATICA; the $\mathtt{R}$ code is available from the authors upon request.

\section{Efficient goodness-of-fit tests for (R)IG distributions within the GIG model.} \label{sec:IG}

The  most important sub-model nested by the GIG family is the {inverse Gaussian}  distribution, obtained for $p=-\frac{1}{2}$. The IG distribution with its two positive parameters $a$ and $b$ has been introduced to the statistical community by Tweedie (1945, 1956, 1957) and has since been revealed to be quite convenient in modeling and analyzing  observations that are  right-skewed and positive (see Seshadri~1999). Among the numerous domains of application figure  fields such as cardiology, demography,
finance,  hydrology or pharmacokinetics. For  references which give illustrative
applications of the IG distribution, we quote Chhikara and
Folks~(1989), Seshadri~(1993) as well as Seshadri~(1999). 
One further reason for the popularity of the IG  are the many similarities, in terms of statistical properties, between the Gaussian and the inverse Gaussian families, as pointed out \mbox{e.g.} by Mudholkar and Tian (2002). As its reciprocal distribution, the RIG of course benefits from this popularity and has also awakened the interest of the statistical community (see, e.g., Scaillet~2004).

In view of its relevance, we particularize now our findings from the previous section to $p_0=-1/2$, the IG sub-model. Thanks to the previously stated property of GIG laws,  ${\rm RIG}(b,a)$ is the law of $1/X$ if  $X \sim {\rm
IG}(a, b)$, hence we shall obtain  for free as well a test for $p_0=1/2$. Under the IG model, the MM estimates of the parameters $a$ and $b$ admit the nice expressions   
$\hat{a}\n_{\rm MM}= \frac{1}{\bar{X}(\bar{X}\bar{X}_{-1}-1)}    $
 and $\hat{b}\n_{\rm MM}=  \frac{\bar{X}}{\bar{X}\bar{X}_{-1}-1}, $ which happen to correspond with the ML estimates (see Seshadri 1999, page 7, a result dating back to Tweedie 1957). Further simplifications arise in the Fisher information matrix ${\it \Gamma} (\varthetab)$ as 
$$
\Gamma_{a,b}(-1/2,a,b)=\frac{-1}{4\sqrt{ab}}, \,\, \Gamma_{a,a}(-1/2,a,b)=\frac{\sqrt{b}a^{-3/2}}{4} \,\,\mbox{and}\,\,
\Gamma_{b,b}(-1/2,a,b)=\frac{1}{2b^2}+ \frac{\sqrt{a}b^{-3/2}}{4}.$$ 
Next, denote by $u$, $v$ and $w$, respectively, the values of $\partial_{pp}^2 \log K_p(\sqrt{ab})$, $\partial_{pa}^2 \log K_p(\sqrt{ab})$ and $\partial_{pb}^2 \log K_p(\sqrt{ab})$ at $p=-1/2$. We then have 
\begin{equation*}
{\it \Gamma}(-1/2,a,b)= \left( \begin{array}{ccc}
u& v-\frac{1}{2a} & w+\frac{1}{2b}\\
   v-\frac{1}{2a} & \frac{\sqrt{b}a^{-3/2}}{4}  &
   -\frac{1}{4\sqrt{ab}}  \\
 w+\frac{1}{2b}   & -\frac{1}{4\sqrt{ab}} & 
 \frac{1}{2b^2}+ \frac{\sqrt{a}b^{-3/2}}{4}
\end{array}\right),
\end{equation*}
yielding $\Gamma_{a,a}(-1/2,a,b)\Gamma_{b,b}(-1/2,a,b)-(\Gamma_{a,b}(-1/2,a,b))^2=\frac{1}{8(ab)^{3/2}}$ and
\begin{equation*}
\Gamma^{eff}_{p,p}(-1/2,a,b)= u-
(2av-1)^2 \left(\frac{1}{\sqrt{ab}}+\frac{1}{2}\right)- (2av-1)(2bw+1) -\frac{1}{2} (2bw+1)^2.
\end{equation*}
The expression of $\Gamma^{eff}_{p,p}(-1/2,a,b)$ can be  calculated efficiently   using the integral expressions  of the entries of ${\it \Gamma} (\varthetab)$ given at the end of the previous section.

Regarding the efficient central sequence, we find
 \begin{equation*}
 \Delta^{(n)eff}_p(-1/2,a,b)= \frac{1}{\sqrt{n}}\sum_{i=1}^n \left( \log(X_i) - \frac{M_{-3/2}}{I_{-3/2}} + R_i(a,b) \right),       
\end{equation*} 
 with
 \begin{eqnarray*}
 R_i(a,b)& =& X_i \left[ v\left(2a\sqrt{\frac{a}{b}}+a^2 \right) + w ab - \sqrt{\frac{a}{b}} \right]+
 \frac{1}{X_i}(wb^2+vab) \nonumber \\
 & & -va(3+2\sqrt{ab})- wb(1+2\sqrt{ab})+1.\nonumber
 \end{eqnarray*}
 Quite interestingly, if $a$ and $b$ are replaced with their MM (or ML) estimates, one gets $\sum_{i=1}^n R_i(\hat{a}^{(n)}_{\rm MM},\hat{b}^{(n)}_{\rm MM}) =0$, so that 
  \begin{equation*}
 \Delta^{(n)eff}_p (-1/2,\hat a\n_{\rm MM},\hat b\n_{\rm MM})= \frac{1}{\sqrt{n}}\sum_{i=1}^n \left( \log(X_i) - \frac{\hat{M}_{-3/2}}{\hat{I}_{-3/2}}\right),       
\end{equation*}
where $\hat{I}_{-3/2}=I_{-3/2}(\hat{a}^{(n)}_{\rm MM},\hat{b}^{(n)}_{\rm MM})$ and $\hat{M}_{-3/2}=M_{-3/2}(\hat{a}^{(n)}_{\rm MM},\hat{b}^{(n)}_{\rm MM})$.   Summing up, the optimal goodness-of-fit test $\phi^{(n)}_{-1/2}$ rejects at asymptotic level $\alpha$ the null hypothesis of an ${\rm IG}(a,b)$ model in favor of a general ${\rm GIG}(p,a,b)$ model with $p\neq-1/2$ whenever the test statistic
\begin{equation*}
|Q^{(n)}_{-1/2}|:=\frac{\left|\frac{1}{\sqrt{n}}\sum_{i=1}^n \left( \log(X_i) - \frac{\hat{M}_{-3/2}}{\hat{I}_{-3/2}}\right)\right|}{\sqrt{\Gamma^{eff}_{p,p}(-1/2,\hat{a}^{(n)}_{\rm MM},\hat{b}^{(n)}_{\rm MM})}}
\end{equation*}
exceeds $z_{\alpha/2}$.

\section{Monte Carlo simulation study.}\label{sec:MC}

In this section we study the finite-sample behavior of  our testing  procedures. More precisely, we focus on the goodness-of-fit test for IG models described explicitly in Section~\ref{sec:IG}, and we assess its performances by comparing our optimal test $\phi^{(n)}_{-1/2}$  to the likelihood ratio test $\phi^{(n)}_{\rm LR}$. The Monte Carlo simulation study has been conducted on~$\mathtt{R}$.

We start by investigating in how far the level constraint is met under the null hypothesis. To this end, we have generated $N=10,000$ independent samples of varying small sample sizes $n=100, 50$ and even $30$ and for distinct values of the parameters $a$ and $b$, and run both tests $\phi^{(n)}_{-1/2}$ and $\phi^{(n)}_{\rm LR}$ at the $\alpha=5\%$ level. The results are summarized in Table~\ref{tab1}. We see that for each choice of couple $(a,b)$, the level is well maintained  by our test when $n=100$ and $n=50$ (although, of course, less well for $50$ than for $100$ observations). Regarding  the very small sample size $n=30$, it still  gives rise to  acceptable levels, although some values are very low. This shows that, even for (very) small sample sizes, $\phi^{(n)}_{-1/2}$ meets the nominal level constraint, especially when compared to the likelihood ratio test $\phi^{(n)}_{\rm LR}$ which attains the level not as well as our novel test.

\begin{table} 
\begin{center}
\begin{tabular}{|l|ccc|ccc|ccc|}
\hline
Size & \multicolumn{3}{c}{$n=100$}\vline&  \multicolumn{3}{c}{$n=50$}\vline& \multicolumn{3}{c}{$n=30$}\vline \\ \hline
 $(a,b)$& $(1,5)$& $(3,3)$&$(3,2)$&$(1,5)$&$(3,3)$&$(3,2)$& $(1,5)$& $(3,3)$&$(3,2)$\\\hline
$\phi^{(n)}_{-1/2}$ &0.0443  &  0.0440 & 0.0452 &  0.0407       & 0.0407   &0.0414  &	0.0370	  & 0.0350&0.0368  \\
$\phi^{(n)}_{\rm LR}$& 0.0403& 0.0388 &0.0398&0.0294&0.0278&0.0314&0.0187&0.0162&0.0171\\
\hline
\end{tabular}
\caption{Rejection frequencies (out of $N=10,000$ replications) of our optimal goodness-of-fit test $\phi_{-1/2}^{(n)}$ and of the likelihood ratio test $\phi^{(n)}_{\rm LR}$ under three distinct null hypotheses ${\rm IG}(1,5)$, ${\rm IG}(2,2)$ and ${\rm IG}(3,2)$  and, in each setting, for three different sample sizes $n=100, 50$ and $30$. The nominal level is $\alpha=0.05$.} \label{tab1}
\end{center}
\end{table}

Now, the level constraint being checked, we compare the power of our test $\phi^{(n)}_{-1/2}$ to the likelihood ratio test $\phi^{(n)}_{\rm LR}$ for small and moderate sample sizes $n=100$ and $n=200$ and for distinct combinations of parameters $(a,b)$. To this end, we have once more generated $N=10,000$ independent samples, this time with $p=-\frac{1}{2} +(\delta-5) \frac{1}{2}$ for $\delta=1,\ldots,9$. This choice permits us to test the power of our test against both higher and lower values of the parameter of interest. Note that, for $\delta=6$, we test against the hyperbolic distribution, whereas for $\delta=7$ the alternative is of RIG type. The results at the $5\%$ level are reported in Table~\ref{tab2}. We clearly see that the two tests detect the deviation from the null hypothesis for any combination $(a,b)$, the performance increasing logically with the sample size. The differences in performance are quite small, but the more significant differences are always in favor of our test, which is yet another argument that our simple efficient test supersedes the classical test $\phi^{(n)}_{\rm LR}$.

\begin{table} 
\begin{tabular}{|l|ccccccccc|}
\hline
& $\delta=1$& $\delta=2$&$\delta=3$&$\delta=4$&$\delta=5$&$\delta=6$& $\delta=7$& $\delta=8$&$\delta=9$\\\hline
& \multicolumn{9}{c}{$(a,b)=(1,2)$ and $n=100$}\vline \\ \hline
$\phi^{(n)}_{-1/2}$ &0.3252 &0.2741& 0.1697 &0.0781 &0.0444& 0.0910& 0.2206& 0.3742 &0.4926 \\
$\phi^{(n)}_{\rm LR}$ & 0.3006 & 0.2505 & 0.1548 & 0.0688 & 0.0406 & 0.0921 & 0.2227 & 0.3762 & 0.4924\\
\hline
& \multicolumn{9}{c}{$(a,b)=(2,2)$ and $n=100$}\vline \\ \hline
$\phi^{(n)}_{-1/2}$ &0.2213 &0.1754& 0.1238 &0.0635 & 0.0429& 0.0684& 0.1264& 0.2141& 0.2972 \\
$\phi^{(n)}_{\rm LR}$ & 0.2004 & 0.1562 & 0.1116 & 0.0560 & 0.0389 & 0.0668 & 0.1246 & 0.2117 & 0.2933\\
\hline
& \multicolumn{9}{c}{$(a,b)=(4,2)$ and $n=100$}\vline \\ \hline
$\phi^{(n)}_{-1/2}$ &0.1513& 0.1094 & 0.0811 &0.0553 &0.0461 & 0.0541 &0.0810 &0.1190 &0.1716 \\
$\phi^{(n)}_{\rm LR}$ & 0.1330 & 0.0974 & 0.0710 & 0.0476 & 0.0404 & 0.0516 & 0.0759 & 0.1148 & 0.1664\\
\hline
& \multicolumn{9}{c}{$(a,b)=(1,2)$ and $n=200$}\vline \\ \hline
$\phi^{(n)}_{-1/2}$ &0.5757& 0.4747 &0.3051 &0.1271 &0.0458& 0.1377 &0.3926& 0.6509& 0.7872 \\
$\phi^{(n)}_{\rm LR}$& 0.5613 & 0.4584 & 0.2907 & 0.1190 & 0.0435 & 0.1400 & 0.3988 & 0.6538 & 0.7895\\
\hline
& \multicolumn{9}{c}{$(a,b)=(2,2)$ and $n=200$}\vline \\ \hline
$\phi^{(n)}_{-1/2}$ & 0.4098 &0.3007 &0.1833 &0.0885 &0.0484 &0.0917 &0.2134 &0.3910 &0.5333 \\
$\phi^{(n)}_{\rm LR}$ & 0.3928 & 0.2878 & 0.1714 & 0.0835 & 0.0474 & 0.0923 & 0.2151 & 0.3924 & 0.5339\\
\hline
& \multicolumn{9}{c}{$(a,b)=(4,2)$ and $n=200$}\vline \\ \hline
$\phi^{(n)}_{-1/2}$ &0.2568& 0.1860 &0.1178& 0.0684& 0.0486& 0.0684& 0.1265 &0.2140 &0.3088 \\
$\phi^{(n)}_{\rm LR}$ & 0.2426 & 0.1739 & 0.1095 & 0.0634 & 0.0460 & 0.0673 & 0.1254 & 0.2120 & 0.3068\\
\hline
\end{tabular}
\begin{center}
\caption{Rejection frequencies (out of $N=10,000$ replications) of our optimal test $\phi_{-1/2}^{(n)}$ and the likelihood ratio test $\phi^{(n)}_{\rm LR}$ under ${\rm GIG}(-\frac{1}{2}+(\delta-5)\frac{1}{2},a,b)$ distributions for $\delta=1,2,\ldots,9$ and several choices of $(a,b)$  and, in each setting, for two different sample sizes $n=100$ and $200$. The nominal level is $\alpha=0.05$.} \label{tab2}
\end{center}
\end{table}


\section{Real-data example.}\label{sec:real}

In this section, we analyze two real-data examples in the light of the new test $\phi^{(n)}_{-1/2}$ developed in this paper. 

Our first example concerns the traffic data example analyzed in Section 7.3 from J\o rgensen~(1982). It concerns the length (in seconds) of $n=128$ intervals between the times at which vehicles pass a point on a road; these values correspond to 2.8, 3.4, 1.4, 14.5, 1.9, 2.8, 2.3, 15.3, 1.8, 9.5, 2.5, 9.4, 1.1, 88.6, 1.6, 1.9, 1.5, 33.7, 2.6, 12.9, 16.2, 1.9, 20.3, 36.8, 40.1, 70.5, 2, 8, 2.1, 3.2, 1.7, 56.5, 23.7, 2.4, 21.4, 5.1, 7.9, 20.1, 14.9, 5.6, 51.7, 87.1, 1.2, 2.7, 1, 1.5, 1.3, 24.7, 72.6, 119.8, 1.2, 6.9, 3.9, 1.6, 3, 1.8, 44.8, 5, 3.9, 125.3, 22.8, 1.9, 15.9, 6, 20.6, 12.9, 3.9, 13, 6.9, 2.5, 12.3, 5.7, 11.3, 2.5, 1.6, 7.6, 2.3, 6.1, 2.1, 34.7, 15.4, 4.6, 55.7, 2.2, 6, 1.8, 1.9, 1.8, 42, 9.3, 91.7, 2.4, 30.6, 1.2, 8.8, 6.6, 49.8, 58.1, 1.9, 2.9, 0.5, 1.2, 31, 11.9, 0.8, 1.2, 0.8, 4.7, 8.3, 7.3, 8.8, 1.8, 3.1, 0.8, 34.1, 3, 2.6, 3.7, 41.3, 29.7, 17.6, 1.9, 13.8, 40.2, 10.1, 11.9, 11 and 0.2. We want to test whether the IG and RIG models fit well the data within the GIG family. For the null hypothesis of an IG model, our test $\phi^{(n)}_{-1/2}$, performed under two-sided form, gives the p-value 0.0642, leading to a light non-rejection of the IG model, whereas the null of an RIG model is heavily rejected with a p-value of $1.8232\times 10^{-5}$. We thus conclude that the latter model does not fit the data, whereas the IG model can serve as candidate inside the GIG family. Note that this finding is not in contradiction with the result of Natarajan and Mudholkar~(2004) whose test clearly rejects the IG model for this data set, because they do not test IG within the GIG family but against other types of alternatives.

Our second example deals with the repair time data of Section 7.4 in J\o rgensen (1982). The data read 
0.2, 0.3, 0.5, 0.5, 0.5, 0.5, 0.6, 0.6, 0.7, 0.7, 0.7, 0.8, 0.8, 1, 1, 1, 1, 1.1, 1.3, 1.5, 1.5, 1.5, 1.5, 2, 2, 2.2, 2.5, 2.7, 3, 3, 3.3, 3.3, 4, 4, 4.5, 4.7, 5, 5.4, 5.4, 7, 7.5, 8.8, 9, 10.3, 22 and 24.5 and correspond to $n=46$ active repair times (in hours) for an airborne communication transceiver. J\o rgensen~(1982) has concluded on basis of a Kolmogorov-Smirnov test that the IG model is a good underlying distribution for the repair times. Our test $\phi^{(n)}_{-1/2}$ completely agrees with his analysis, with a p-value of 0.4484, whereas the RIG model is here as well rejected (at the $5\%$ level) with a p-value of 0.0207.

\section{Future research.}\label{sec:future}

Yet another interesting aspect of the Le Cam methodology consists in the fact that it allows for constructing a simple efficient estimator for $\varthetab=(p,a,b)'$, the so-called one-step estimator. The main idea behind one-step estimation consists in adding to an existing adequate preliminary estimator $\hat{\varthetab}^{(n)}$ a quantity depending on a version of the  central sequence for $\varthetab$. More precisely, the one-step estimator takes on the guise
\begin{align*} \hat{\varthetab}_{Cam}^{(n)}=\hat{\varthetab}^{(n)}+ n^{-1/2}\left({\pmb \Gamma} (\hat\varthetab\n)\right)^{-1}\Deltab^{(n)}(\hat\varthetab\n),\end{align*}
where $\hat\varthetab\n=(\hat p\n,\hat a\n,\hat b\n)'$ is a preliminary estimator of $\varthetab$ fulfilling Assumption~A (we would of course choose the MM estimator). The resulting estimator combines simplicity with efficiency, as, like the test statistics developed in this paper, it allows for a closed-form expression but is as efficient as ML estimators. Establishing the asymptotic behavior of such estimators and their optimality features, as well as comparing their performance to the competitors of the literature in a detailed simulation study, is part of ongoing research.

%

 \vspace{0.5cm}

\noindent ACKNOWLEDGEMENTS\vspace{2mm}

\noindent The research of Christophe Ley is supported by a Mandat de Charg\'e de Recherche FNRS from the Fonds National de la Recherche Scientifique, Communaut\'e fran\c caise de Belgique. The authors also thank Ivan Nourdin for inviting Christophe Ley for a visit at the Institut Elie Cartan, research stay during which the present work was initiated and the main parts worked out. 
\vspace{0.5cm}



\appendix

\section{Appendix.}

\noindent {\sc Proof of Theorem~\ref{ULAN}.} Establishing the ULAN property of ${\rm GIG}(p,a,b)$ with respect to all three parameters $p,a,b$ is quite straightforward since we are not working within a semiparametric family of distributions (hence we do not have to deal with an infinite-dimensional parameter); the problem considered involves a parametric family of distributions with densities meeting the most classical regularity conditions.
 In particular, one readily obtains that (i) $(p,a,b)\mapsto \sqrt{c(p,a,b) x^{p-1}e^{-(ax+b/x)/2}}$ is continuously differentiable for every $x>0$ and (ii) the associated Fisher information matrix is well defined and continuous in $p,a$ and $b$. Thus, by Lemma 7.6 of van der Vaart~(1998), $(p,a,b)\mapsto \sqrt{c(p,a,b) x^{p-1}e^{-(ax+b/x)/2}}$ is differentiable in quadratic mean, and the ULAN property follows from Theorem 7.2 of van der Vaart~(1998). This completes the proof. \hfill \cqfd
\vspace{1cm}

\noindent{\sc Proof of Proposition~\ref{aslinprop}.}
It follows from the asymptotic linearity property of the central sequence given in (\ref{aslin}) that, under  ${\rm P}_{\varthetab}^{(n)}$ and for $n\rightarrow\infty$,
\begin{align*}
{\Deltab}^{(n)eff}_p\left(\varthetab+n^{-1/2} (0,\tau_2^{(n)},\tau_3^{(n)})'\right)={\Deltab}^{(n)eff}_p(\varthetab)-{\it P}(\varthetab){\it \Gamma}(\varthetab) \left(
\begin{array}{c}
0\\
\tau_2^{(n)}\\
\tau_3^{(n)}
\end{array}
\right)+o_{\rm{P}}(1),
\end{align*}
where 
$$
{\it P}(\varthetab)=\left(1, -(\Gamma_{p,a}(\varthetab),\Gamma_{p,b}(\varthetab))\left(\begin{array}{cc}
\Gamma_{b,b}(\varthetab)&-\Gamma_{a,b}(\varthetab)\\
-\Gamma_{a,b}(\varthetab)&\Gamma_{a,a}(\varthetab)
\end{array}\right)/(\Gamma_{b,b}(\varthetab)\Gamma_{a,a}(\varthetab)-(\Gamma_{a,b}(\varthetab))^2)\right)
$$
is the projection matrix onto the subspace orthogonal to the central sequences for~$a$ and~$b$. One easily checks that the  product ${\it P}(\varthetab){\it \Gamma}(\varthetab)$ is of the form $(\cdot,0,0)$, leading to the announced asymptotic linearity~(\ref{aslineff}). The other asymptotic equality in probability follows by continuity.
 \cqfd \vspace{1cm}

\noindent{\sc Proof of Theorem~\ref{behav}.}
\noindent The statement in Part (i) is easily proved thanks to the asymptotic linearity in~(\ref{replace}) under the null, since $\Delta^{(n)eff}_p(p_0,a,b)$ is asymptotically $\mathcal{N}(0, \Gamma_{p,p}^{eff}(p_0,a,b))$ under $\bigcup_{a\in\R_0^+} \bigcup_{b\in\R_0^+}{\rm P}^{(n)}_{p_0,a,b}$ by the central limit theorem. 

In order to prove the more delicate Part~(ii), observe that,  under ${\rm P}^{(n)}_{p_0,a,b}$ and for any bounded sequence ${\taub}^{(n)}=(\tau_1^{(n)}, {\tau}_2^{(n)}, \tau_3^{(n)}) '\in\R^3$, we see that, as $n\rightarrow\infty$, 
$$\begin{pmatrix} 
\Delta_p^{(n)eff}(p_0,a,b) \\
\Lambda^{(n)}_{(p_0,a,b)'+n^{-1/2}\taub^{(n)}/(p_0,a,b)'}
\end{pmatrix} \stackrel{\mathcal{L}}{\longrightarrow}\mathcal{N}_2 \left( \begin{pmatrix} 0 \\ -\frac{1}{2} \taub'{\it \Gamma}(\varthetab)\taub\end{pmatrix}, \begin{pmatrix} \Gamma_{p,p}^{eff}(p_0,a,b) & \tau_1\Gamma_{p,p}^{eff}(p_0,a,b)\\
\tau_1 \Gamma_{p,p}^{eff}(p_0,a,b) &  \taub'{\it \Gamma}(\varthetab)\tau\end{pmatrix}\right),$$
where $\Lambda^{(n)}_{(p_0,a,b)'+n^{-1/2}\taub^{(n)}/(p_0,a,b)'}$ is the log-likelihood ratio and  $\taub=\lim_{n \to \infty} \taub^{(n)}$. We can then apply Le Cam's third lemma which implies that $\Delta_{p}^{(n)eff}(p_0,a,b)$ is asymptotically
$\mathcal{N}\left(\tau_1\Gamma_{p,p}^{eff}(p_0,a, b), \Gamma_{p,p}^{eff}(p_0,a,b)\right)$ under ${\rm P}^{(n)}_{p_0+n^{-1/2}\tau_1^{(n)},a,b}$. Since the asymptotic linearity (\ref{replace}) holds as well under ${\rm P}^{(n)}_{p_0+n^{-1/2}\tau_1^{(n)},a,b}$ by contiguity, Part (ii) of the theorem readily follows.

As regards Part (iii), the fact that $\phi^{(n)}_{p_0}$ has  asymptotic level $\alpha$ follows directly from the asymptotic null distribution given in Part (i), while local asymptotic maximinity is a consequence of the  convergence of the local experiments to the Gaussian shift experiment (see Le Cam and Yang~2000). \cqfd \vspace{1cm}

\noindent {\bf References}
\begin{enumerate}
\item[] Barndorff-Nielsen, O. E. and Halgreen, C. (1977). Infinite divisibility of the Hyperbolic and generalized inverse Gaussian distribution. {\it Z. Wahrscheinlichkeitstheorie und Verw. Gebiete}   {\bf 38}, 309--312.

\item[]
Chebana, F., El Adlouni, S. and Bob\'ee, B. (2010). Mixed estimation methods for Halphen distributions  with applications in extreme hydrologic events. \emph{Stoch. Environ. Res. Risk Assess.}~{\bf 24}, 359--376.

\item[]
Chhikara, R. S. and  Folks, L. (1989). \emph{The inverse Gaussian distribution: theory,
methodology, and applications}. Marcel Dekker, New York.


\item[]
Fitzgerald, D. L. (2000). Statistical aspects of Tricomi's function and
modified Bessel functions of the second kind. \emph{Stoch. Environ. Res. Risk Assess.}~{\bf 14}, 139--158.

\item[]
Good, I. J. (1953). The population frequencies of species and the estimation of population parameters. \emph{Biometrika}~{\bf 40}, 237--264.

\item[]
Halphen, E. (1941). Sur un nouveau type de courbe de fr\'equence. \emph{Comptes Rendus de l'Acad\'emie des Sciences}~{\bf 213}, 633--635. Published under the name of ``Dugu\'e'' due to war constraints.

\item[]
Iyengar, S. and Liao, Q. (1997). Modeling neural activity using the generalized inverse Gaussian distribution. \emph{Biol. Cybern.}~{\bf 77}, 289--295.

\item[]
J\o rgensen B. (1982). \emph{Statistical properties of the generalized inverse Gaussian distribution}. Springer-Verlag, Heidelberg.

\item[]
Kreiss, J-P. (1987). On adaptive estimation in stationary ARMA processes. \emph{Ann. Statist.} {\bf 15}, 112--133.

\item[] 
 Le Cam, L. (1960). Locally asymptotically normal families of distribution. \emph{Univ. Calif. Publ. in Stat.} \textbf{3}, 37--98.

 \item[] Le Cam, L. and Yang, G. L. (2000). \emph{Asymptotics in statistics. Some basic concepts}. 2nd ed. Springer-Verlag, New York.

\item[]
Lemonte, A. J. and Cordeiro, G. M. (2011). The exponentiated generalized inverse Gaussian distribution. \emph{Stat. Probab. Lett.}~{\bf 81}, 506--517.

\item[] Mudholkar, S. M. and Tian, L. (2002). An entropy characterization of the inverse Gaussian distribution and related goodness-of-fit test. \emph{J. Stat. Plan. Infer.}~{\bf 102}, 211--221.

\item[]
Natarajan, R. and Mudholkar, G. S. (2004). Moment-based goodness-of-fit tests for the inverse Gaussian distribution. \emph{Technometrics}~{\bf 46}, 339--347.

\item[]
Perreault, L., Bob\'ee, B. and Rasmussen, P. F. (1999a). Halphen distribution system. I: Mathematical and statistical properties. \emph{J. Hydrol. Eng.}~{\bf 4}, 189--199.

\item[]
Perreault, L., Bob\'ee, B. and Rasmussen, P. F. (1999b). Halphen distribution system. II: Parameter and quantile estimation. \emph{J. Hydrol. Eng.}~{\bf 4}, 200--208.

\item[]
Scaillet, O. (2004). Density estimation  using inverse and reciprocal inverse Gaussian kernels. \emph{J. Nonparametr. Stat.}~{\bf 16}, 217--226.

\item[]
Seshadri, V. (1993). \emph{The inverse Gaussian distribution--a case study in exponential families}. Oxford University Press.

\item[]
Seshadri, V. (1999). \emph{The inverse Gaussian distribution}. Springer-Verlag, New York.



\item[]
Tweedie, M. C. K. (1945). Inverse statistical variates. \emph{Nature}~{\bf 155}, 453--453.

\item[]
Tweedie, M. C. K. (1956). Statistical properties of inverse gaussian distributions. \emph{Virginia Journal of Science}~{\bf 7}, 160--165.

\item[]
Tweedie, M. C. K. (1957). Statistical properties of inverse gaussian distributions. \emph{Ann.  Math. Statist.}~{\bf 28}, 362--377.

\item[]
van der Vaart, A. W. (1998). \emph{Asymptotic Statistics}. Cambridge Series in Statistical
and Probabilistic Mathematics, Cambridge University Press.


\end{enumerate}

\end{document}